\documentclass[%
reprint,
amsmath,amssymb,
aps,
]{revtex4-1}
\usepackage{graphicx}
\usepackage{hyperref}

\begin{document}

\title{Impurity effect as a probe for the pairing symmetry of graphene-based superconductors}

\author{Yuan-Qiao Li$^{1}$}
\author{Tao Zhou$^{2,3}$}%
\email{Corresponding author: tzhou@scnu.edu.cn}
\affiliation{$^{1}$College of Science, Nanjing University of Aeronautics and Astronautics, Nanjing 210016, China\\
$^2$Guangdong Provincial Key Laboratory of Quantum Engineering and Quantum Materials, School of Physics and Telecommunication Engineering, South China Normal University, Guangzhou 510006, China\\
$^3$Guangdong-Hong Kong Joint Laboratory of Quantum Matter, Frontier Research Institute for Physics, South China Normal University, Guangzhou 510006, China
}
\date{\today}

\begin{abstract}
The single impurity effect on the graphene-based superconductor is studied theoretically. Four different pairing symmetries are discussed.
Sharp resonance peaks are found near the impurity site for the $d+id$-wave pairing symmetry and the $p+ip$-wave pairing symmetry
when the chemical potential is large. As the chemical potential decreases, the in-gap states are robust for the $d+id$ pairing symmetry while they disappear for the $p+ip$ pairing symmetry. Such in-gap peaks are absent for the fully gapped
extended $s$-wave pairing symmetry and the nodal $f$-wave pairing symmetry. The existence of the in-gap resonance peaks can be explained
well based on the sign-reversal of the superconducting gap along different Fermi pockets and by analyzing the denominator of the $T$-matrix. All of the features can be accessed by the experiments, which provide a useful probe for
the pairing symmetry of graphene-based superconductors.
\end{abstract}
\maketitle

\section{introduction}
The graphene-based superconductors have attracted great interest in the past few years.
Experimentally, superconductivity was proposed to be induced to the monolayer graphene growing
on the superconducting Rhenium film~\cite{tonn}. Evidence of superconductivity was also observed
on a Ca-intercalated bilayer graphene~\cite{chin,chap}, and Li-decorated monolayer graphene~\cite{ludb}.
And superconductivity was also reported to be induced to the monolayer graphene by placing it on
an electron-doped cuprate superconductor~\cite{bern}. Quite interestingly, signatures of the $p$-wave pairing
symmetry were observed through the scanning tunnelling spectroscopy (STS) investigation. Recently, it was
reported that superconductivity is realized in the twisted bilayer graphene~\cite{ycao}, which is the
first pure-carbon-based two-dimensional superconductor.

On the theoretical side, the graphene was predicted to go into the superconducting state through doping or
the proximity effect~\cite{ucho,kopn,jacob,anni,tma,faye,nand,kies,nand1,xiao,hoss,lado}. However, the
favored pairing symmetry is still unclear. Different pairing symmetries have been proposed by different
groups. In particular, a $p+ip$-wave symmetry was proposed based on an extended Hubbard model~\cite{tma,faye}.
A $d+id$-wave pairing symmetry was proposed based on the renormalization group method~\cite{nand,kies,nand1} or
the random phase approximation~\cite{xiao}. It was also proposed that the triplet $f$-wave pairing might occur
under some particular conditions~\cite{kies,nand1,xiao}. And an exotic $s$-wave pairing was proposed to be the
favored pairing symmetry in the bilayer graphene~\cite{hoss}. Therefore,
so far there is no agreement about the preferred pairing symmetry in graphene-based superconductors.
Actually, the favored pairing symmetry may depend strongly
on the parameters, the starting model, the pairing mechanism and the approximation considered.
Since identifying the pairing symmetry is crucial to clarify microscopic details of the superconductivity,
it is quite important to provide more detailed experimental information to resolve the pairing symmetry.

The impurity effect has been one powerful tool to explore the pairing symmetry of unconventional superconductors.
One prominent feature is the impurity induced in-gap resonance state in the superconducting state of cuprate
superconductors~\cite{zhu}, this feature was used to identify the $d$-wave pairing symmetry~\cite{jhu,zhu}.
In iron-based superconductors, the existence of the impurity induced in-gap states was proposed to be a signature
of the sign-reversal of the $s_\pm$-wave pairing gap~\cite{zhang,tsai}. Fortunately, STS experiments can directly distinguish various available pairing symmetries
in superconductor and evidence the possible unconventional nature of the superconducting state~\cite{Hudson,Morr,Andrenacci}.
Theoretically, there are two effective methods to investigating the point-like impurity effect in the superconducting state,
the first method is based on the Bogoliubov-de Gennes (BdG) equations~\cite{glad}. The Hamiltonian is diagonalized numerically in the real space. The second method is based on
the self-consistent $T$-matrix method~\cite{zhu,pell,wehl}.
The Hamiltonian is expressed in the momentum space.
Previously, the impurity effects in superconducting graphene have been studied based on these two methods~\cite{pell,wehl,glad}.
With the $T$-matrix method, the impurity effects at a typical doping density were studied~\cite{pell,wehl}.
Recently, based on the BdG technique, the impurity effects of
graphene-based superconductors considering mainly the $d$-wave pairing symmetry were reported~\cite{glad}.
We note that so far a systematic study about the impurity effect in graphene-based unconventional
superconductors based on the $T$-matrix method is still awaited and highly demanded. With the $T$-matrix method, the local density of states (LDOS) is expressed analytically and the origin of the impurity states can be explored numerically.
Another advantage for the $T$-matrix method is that the Hamiltonian is expressed in the momentum space, so that there is no finite-size effect.

Recently, it was reported experimentally that the graphene materials can be highly doped to beyond the Van Hove filling~\cite{Scherer,Rosenzweig}. This experimental progress is of importance and
may open a new door to study the possible superconductivity in graphene-based materials. Therefore,
 it is rather insightful
to study systematically the single impurity effect of the possible superconducting state in the graphene lattice with different doping levels.

In the present work, we study theoretically the impurity effect of a superconductor in the graphene lattice utilizing the $T$-matrix method. Four different pairing
symmetries, i.e., the $p+ip$-wave pairing, the $d+id$-wave pairing, the extended $s$-wave pairing, and the $f$-wave pairing, are considered. Note that, this work is different
from Ref.[26]. To a certain extent, this work will prefect the impurity effect on the different unconventional pairing
symmetries in graphene. For the cases of the extended $s$-wave pairing and the $f$-wave pairing symmetries, no in-gap states are obtained. For the $p+ip$-wave pairing,
the results depend on the doping levels. As the chemical potential is large, there exist sharp resonance peaks for different impurity strengths,
and the resonance peaks disappear as the chemical potential decreases. For the $d+id$-wave pairing symmetry, the impurity induced
in-gap peaks are always lying symmetric with the Fermi energy. As the chemical potential increases, the resonance state at negative
energy is supressed while that at positive energy is enhanced. Our results indicate that the impurity effect may be
useful to probe the pairing symmetry of the graphene-based superconductor.

The rest of the paper is organized as follows.
In Sec. II, we introduce
the model and derive the formalism. In Sec. III, we
report numerical calculations and discuss the obtained
results. Finally, we present a brief summary in Sec. IV.

\section{Model and Hamiltonian}

For the honeycomb lattice which describes the monolayer graphene, each unit cell contains two inequivalent lattice sites. The whole system includes two sublattices $A$ and $B$.
We start from a BCS-type Hamiltonian, including the nearest-neighbor hopping term, the chemical potential term, and the superconducting pairing term.
Then the Hamiltonian is written as,
\begin{equation}\label{H}
H = \sum\limits_{\bf{k}} {\psi _{\bf{k}}^ \dagger  M_{\bf{k}} \psi _{\bf{k}} },
\end{equation}
where
\begin{equation}\label{psi}
\psi _{\bf{k}}^ \dagger   = (A_{{\bf{k}}\uparrow }^\dagger  ,B_{{\bf{k}}\uparrow }^ \dagger  ,A_{ - {\bf{k}} \downarrow }^{} ,B_{ - {\bf{k}} \downarrow }^{} ).
\end{equation}
$A_{{\bf{k}} \uparrow }^ \dagger$ and $B_{{\bf{k}} \uparrow }^ \dagger$ are creation operators of the spin-up electron with momentum $\bf{k}$ on the
sublattices $A$ and $B$, $A_{{\bf{-k}} \downarrow }$ and $B_{{\bf{-k}} \downarrow } $ are annihilation operators of the spin down electron with
momentum $\bf{-k}$ on the sublattices $A$ and $B$, respectively.

$M_{\bf k}$ is the $4\times 4$ matrix in the momentum space:
\begin{equation}\label{mk}
M_{\bf{k}}  = \left( {\begin{array}{*{20}c}
   \vspace{1ex}{-\mu} & {\gamma _{\bf{k}} } & 0 & {\Delta _{\bf{k}} }  \\ \vspace{1ex}
  {\gamma _{\bf{k}} } & { - \mu } & {\xi \Delta _{ - {\bf{k}}} } & 0  \\ \vspace{1ex}
   0 & {\xi \Delta _{ - {\bf{k}}}^* } & {\mu   } & { - \gamma _{\bf{k}} }  \\ \vspace{1ex}
   {\Delta _{\bf{k}} }^* & 0 & { - \gamma _{\bf{k}}^* } & \mu   \\
\end{array}} \right).
\end{equation}
Here $\xi  =  1$ and $-1$ are for the spin triplet pairing and the spin singlet pairing, respectively.
$\gamma _{\bf{k}}$ describes the nearest-neighbor electron hopping, with
\begin{equation}
 \gamma _{\bf{k}}=  - t\sum\limits_{j = 1,2,3} {e^{i{\bf{k}\cdot {\bf e_j} }}},
 \end{equation}
 with  ${\bf{e}}_1=(1,0)$, ${\bf{e}}_2  = \frac{1}{2}( - 1,\sqrt 3 )$, and ${\bf{e}}_3  = \frac{1}{2}( - 1, - \sqrt 3 )$.
 $\Delta _{\bf{k}}$ represents the superconducting pairing. For the electron pairing of the nearest-neighbor sites, it is expressed as,
 \begin{equation}
 \Delta _{\bf{k}}  = \sum\limits_{j = 1,2,3} {\Delta _{j } e^{i{\bf{k}\cdot{\bf e_j }}}}.
 \end{equation}
 Following Ref.~\cite{faye}, for the extended s-wave pairing symmetry and the f-wave pairing symmetry, we have $\Delta _{j }  = \Delta _t (1,1,1)$, and for the $p+ip$-wave pairing  symmetry and the $d+id$-wave pairing symmetry, we have $\Delta _{j }  = \Delta _t (1,e^{i{\textstyle{2 \over 3}}\pi } ,e^{i{\textstyle{4 \over 3}}\pi } )$.

We consider a single impurity being placed on the sublattice $A$ in the unit cell $(0,0)$. The impurity Hamiltonian is written as,
\begin{equation}
H_{imp}=V_s (A^\dagger _{(0,0),\uparrow} A _{(0,0),\uparrow}+ A^\dagger _{(0,0),\downarrow} A _{(0,0),\downarrow}),
\end{equation}
where $V_s$ is the induced imputity strength. Then we can define the $T$-matrix as,
\begin{equation}\label{Tmatrix}
\hat T(\omega ) = {{\hat U_0 } \mathord{\left/
 {\vphantom {{\hat U_0 } {\left[ {\hat I - \hat U_0 \frac{1}{N}\sum\limits_{\bf{k}} {\hat G_0 ({\bf{k}},\omega )} } \right]}}} \right.
 \kern-\nulldelimiterspace} {\left[ {\hat I - \hat U_0 \frac{1}{N}\sum\limits_{\bf{k}} {\hat G_0 ({\bf{k}},\omega )} } \right]}}.
 \end{equation}
 Here $\hat I$ is the $4\times 4$ identity matrix. $\hat G_0 ({\bf{k}},\omega )$ is the bare Green's function in the momentum space, with
 $\hat G_0 ({\bf{k}},\omega )_{ij}=\sum\limits_{n = 1}^4 {\frac{{u_{in} ({\bf{k}})u_{nj}^\dag  ({\bf{k}})}}{{\omega  - E_n ({\bf k})  + i\delta }}}$. $u_{ij} ({\bf{k}})$
 and $E_n ({\bf k})$ are obtained by diagonalizing the
 $4\times4$ Hamiltonian matrix of Eq.\ref{mk}. The non-zero elements of the matrix $\hat U_0$ include $U^{11}_0=V_s$ and $U^{33}_0=-V_s$, respectively.

The local density of states (LDOS) is then expressed as,
\begin{equation}
\rho ({\bf r},{\omega})  =  - \frac{1}{\pi }{\mathop{\rm Im}\nolimits} {\rm{Tr}}\hat G ({\bf{r}},\omega ),
\end{equation}
with
\begin{equation}
\hat G({\bf{r}},\omega ) = \hat G_0 (0,\omega ) + \hat G_0 ({\bf{r}},\omega )\hat T(\omega )\hat G_0 ( - {\bf{r}},\omega ).
\end{equation}
The bare Green's functions $\hat G_0({\bf{r}},\omega )$ in the real space can be obtained by performing a Fourier transformation to the bare
Green's function in the momentum space $\hat G_0 ({\bf{k}},\omega)$.

  \begin{figure}
\centering
  \includegraphics[width=3in]{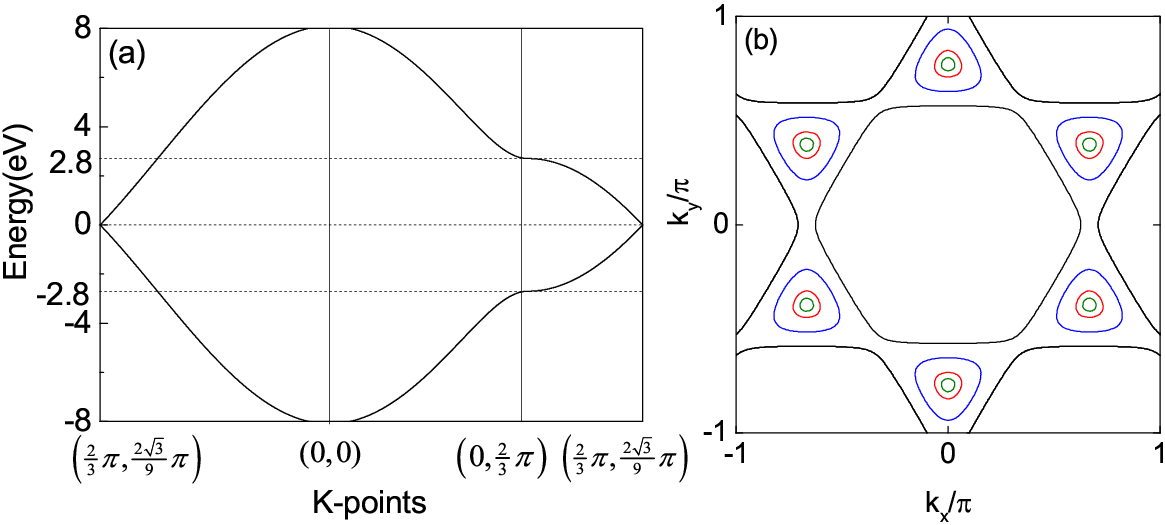}
\caption{(Color online)(a) The normal state energy bands along the high symmetric lines in the Brillouin zone. (b) The normal state Fermi surfaces.
From small to large pockets, the chemical potentials are $0.4$, $0.8$, $1.8$, $2.8$.}
\label{fig1}
\end{figure}

In the results presented below, we use $1$eV as the energy unit. The nearest hopping constant $t$ is chosen as $2.7$.
The gap magnitude is usually very small in real materials. In the present work, as usually done, we consider a much larger
inputting gap to make the detailed in-gap features clear. Our main results do not change qualitatively with different gap magnitudes.

\begin{figure}
\centering
\includegraphics[width=0.4\textwidth]{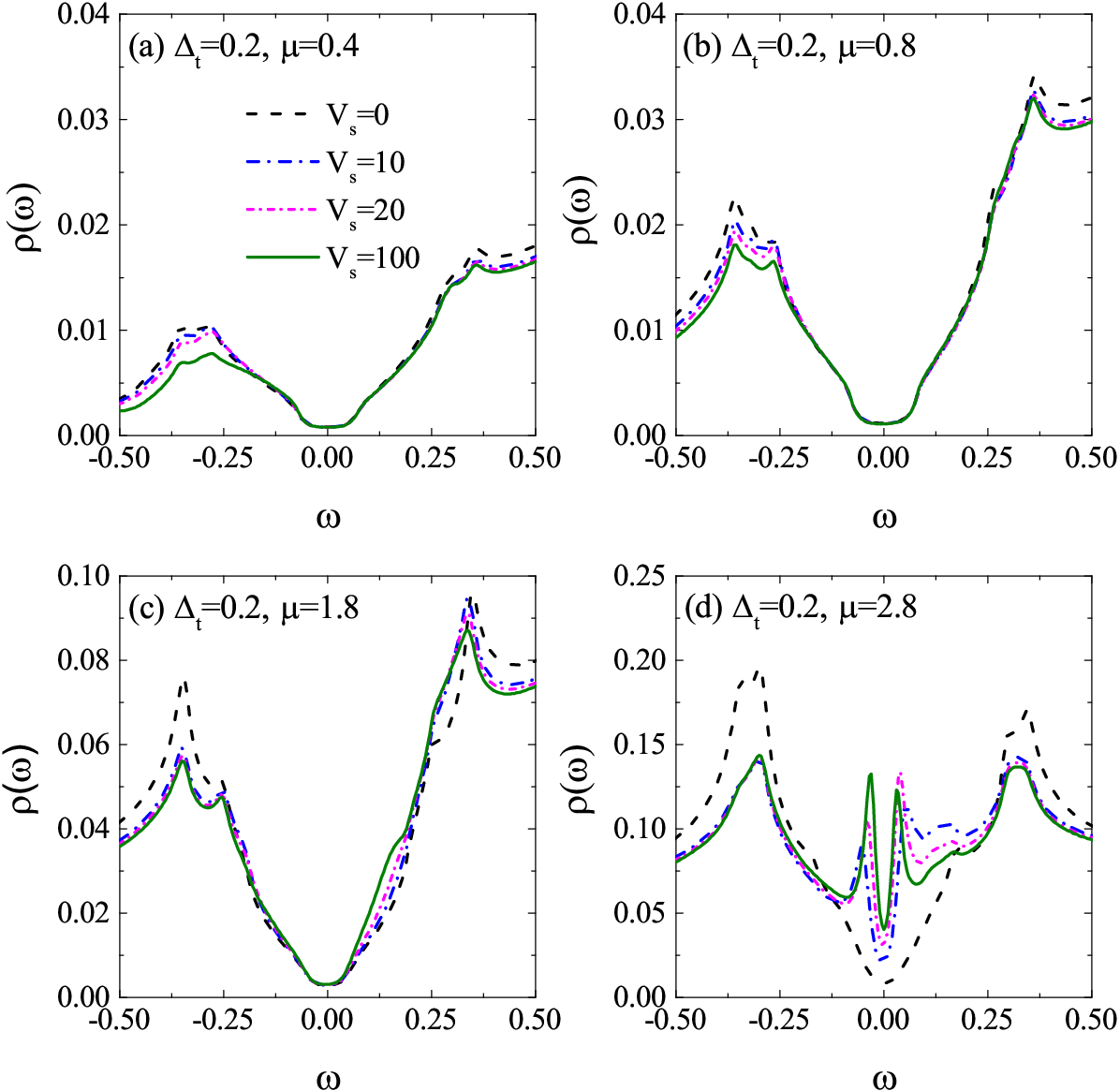}
\caption{(Color online). The LDOS spectra at the nearest-neighbour site of the impurity for the $p+ip$-wave pairing symmetry.}
\label{fig2}
\end{figure}

\section{Results and Discussion}
The normal state energy bands (obtained by setting $\Delta_{\pm\bf k}=0$ in Eq.\ref{mk}) along the high symmetric lines in the Brillouin zone are plotted in Fig.\ref{fig1}(a).
As is seen, the top of the valence band and the bottom of the conduction band touch at the Dirac point. There are two saddle
points at the energies about $\pm 2.8$, which yield the Van Hove singularities in the density of states. As the chemical potential is zero, the Fermi level is located
at the Dirac points. The Fermi surface shrinks to points thus usually the superconductivity cannot occur.
We add a chemical potential term (for doped graphene materials) into the system to pull the Fermi level away from the Dirac points. The Fermi surfaces of the system
for different chemical potentials are shown in Fig.\ref{fig1}(b). As the chemical potential is small, there are six disconnected Fermi pockets.
 The pockets become large when the chemical potential increases. When the Fermi level is doped to near the saddle point, the Fermi pockets connect
 and the Fermi surface becomes a large pocket centered around the Brillouin zone center.

We now study the impurity effect for the $p+ip$-wave pairing symmetry. The LDOS spectra at the nearest-neighbor site of the impurity with different
impurity scattering potentials and chemical potentials are plotted in Fig.\ref{fig2}. For the cases of low doping levels, as is seen in Fig.\ref{fig2}(a-c), no
in-gap resonances exist and the intensities of coherence peaks are supressed as impurity strength increases. The most
prominent feature revealed here is the existence of the in-gap resonance peaks when the Fermi energy reaches the Van Hove singularities, as
presented in Fig.\ref{fig2}(d). At this doping level, two sharp in-gap resonance peaks show up and the resonance peaks locate symmetric with respect to the Fermi energy due to the particle-hole
symmetry of the superconducting Hamiltonian. Note that their strengths are nearly as strong as those of coherence peaks.

\begin{figure}
\centering
\includegraphics[width=0.4\textwidth]{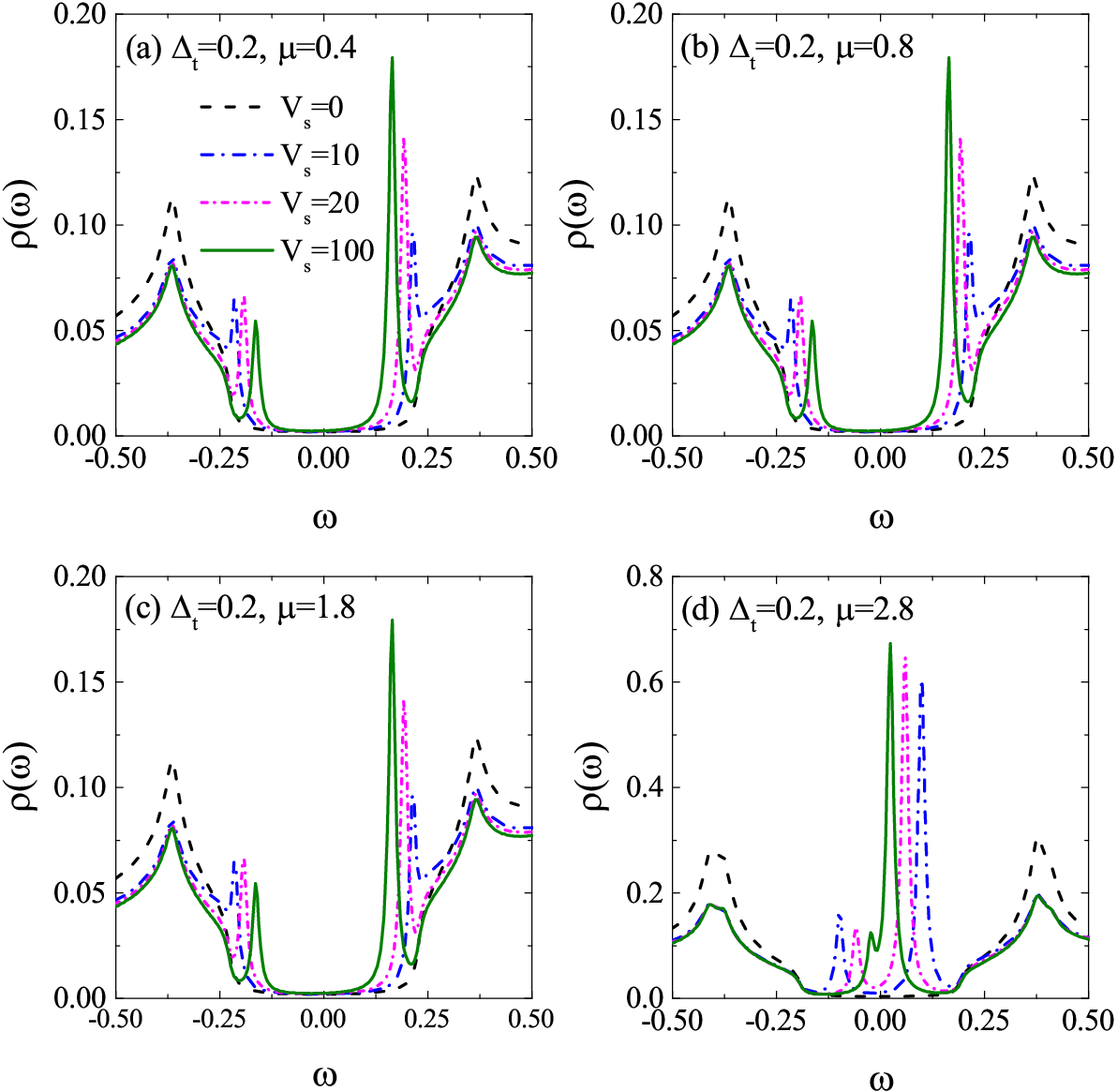}
\caption{(Color online). The LDOS spectra at the nearest-neighbour site of the impurity for the $d+id$-wave pairing symmetry.}
\label{fig3}
\end{figure}

 Let us discuss the impurity states for the $d+id$-wave pairing symmetry. The corresponding LDOS spectra are presented in Fig.\ref{fig3}. Here $U$-shaped spectra are
 obtained for the case of $V_s=0$, indicating that the system is nodeless.
 The energies of the superconducting coherence peaks are larger compared to those of $p+ip$-wave pairing symmetry.
 The larger effective gap indicates that the $d+id$-wave pairing symmetry may be more suitable for the graphene-based energy band, and may have the lower free energy at the
 mean-field level, consistent with previous theoretical predictions~\cite{nand,kies,xiao}. The impurity induced in-gap states for the $d+id$-wave pairing are also
 clearly revealed. As the impurity strength increases, the in-gap states shift to near the Fermi energy. For
 the case of highly doping level ($\mu=2.8$), the intensities of the in-gap peaks are significantly enhanced, qualitatively consistent with previous numerical
 calculations~\cite{lot}. Obviously such strong in-gap features are the resonance states, and their intensities are much larger than that of the superconducting
 coherence peaks, thus it may be easily detected experimentally.

\begin{figure}
\centering
\includegraphics[width=0.4\textwidth]{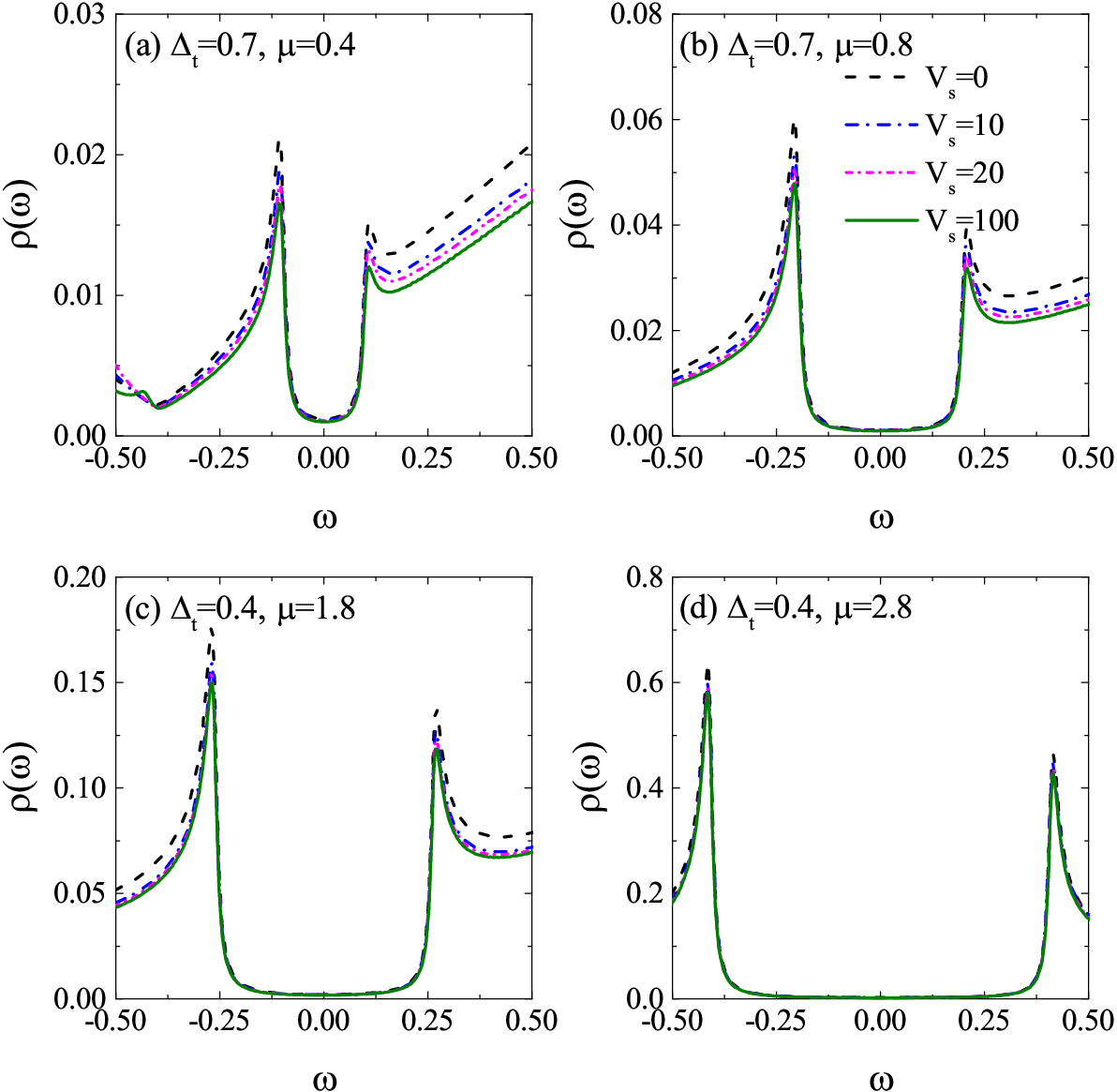}
\caption{(Color online). The LDOS spectra at the nearest-neighbour site of the impurity for the extended $s$-wave pairing symmetry.}
\label{fig4}
\end{figure}

We turn to study the impurity effect for the extended $s$-wave pairing symmetry and the $f$-wave pairing symmetry, respectively. For both the extended $s$-wave pairing
and the $f$-wave pairing, the effective gap magnitude becomes rather small, especially for tiny Fermi pockets when the chemical potential is small.
 Thus we would like to consider a larger input gap magnitude $\Delta_0$ to obtain a large enough effective energy gap.
 In the following presented results, $\Delta_t=0.7$ is used for the cases of $\mu=0.4$ and $0.8$, and $\Delta_t=0.4$ for the cases of $\mu=1.8$ and $\mu=2.8$.

\begin{figure}
\centering
\includegraphics[width=0.4\textwidth]{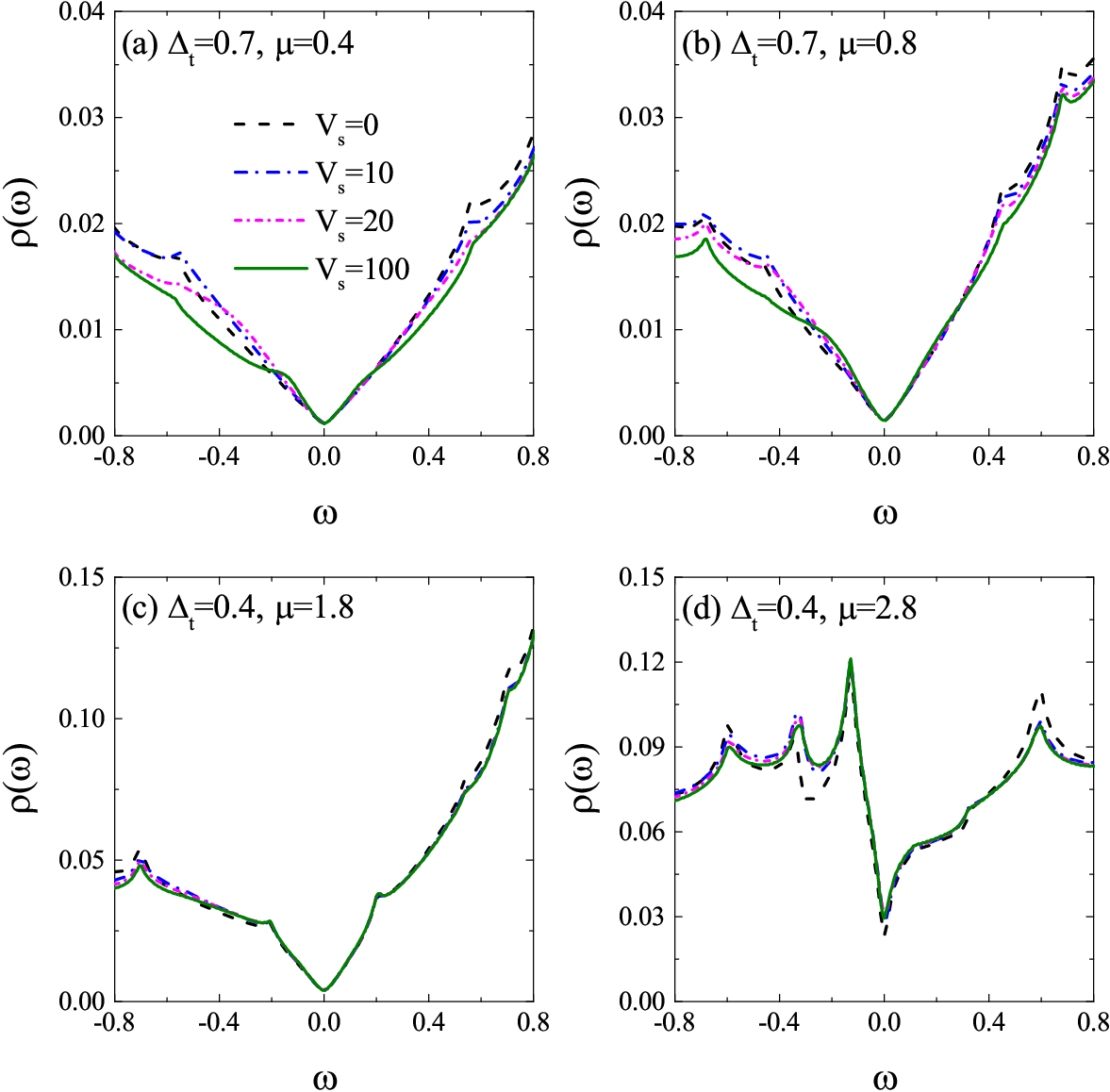}
\caption{(Color online). The LDOS spectra at the nearest-neighbour site of the impurity for the $f$-wave pairing symmetry.}
\label{fig5}
\end{figure}

 The numerical results for the impurity effect in the extended $s$-wave pairing are displayed in Fig.\ref{fig4}. Here the LDOS spectra behave $U$-shaped thus the system
 is fully gapped. As the impurity scattering potential become larger, the superconducting coherence peaks are suppressed due to the impurity scattering. Note that,
 here for all of the parameters we considered, no in-gap resonance peaks exist, which is significantly different from the cases of the $d+id$-wave pairing symmetry.
The numerical results for the cases of the $f$-wave pairing symmetry are presented in Fig.\ref{fig5}. Generally the numerical results are qualitatively the
same with those of extended $s$-wave pairing, except that for the case of $f$-wave pairing, the system is nodal.

 \begin{figure}
\centering
  \includegraphics[width=2.7in]{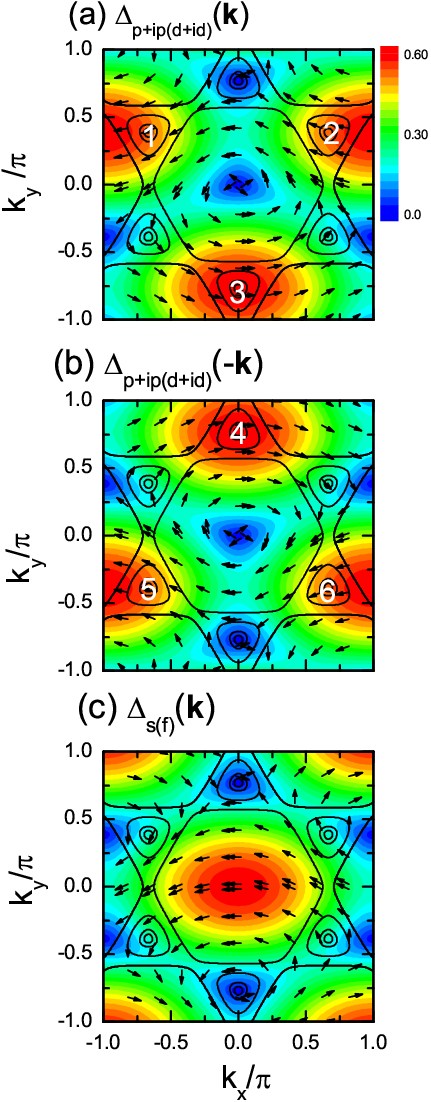}
\caption{(Color online) The intensity plots of superconducting gap magnitude. The arrows indicate their phases. The solid lines are the replots of the normal state Fermi surfaces shown in Fig.~1(b).}
\label{fig6}
\end{figure}

 We here provide a physical picture accounting for the possible in-gap states of the $p+ip$-wave and $d+id$-wave pairing symmetries, through analyzing the superconducting order parameter near the Fermi surface. The superconducting gap magnitudes and their phases [from Eq.(5)] are plotted in Fig.~6.
 As is seen in Figs.~6(a) and 6(b), for the $p+ip$-wave and $d+id$-wave pairing symmetries, the maximum superconducting gap is just near the normal state Fermi pockets, indicated by pockets $1-6$.
The phases of the superconducting gap almost keep the same in one pocket and the phases are reversed for the pockets $1/2$ and $3$ (or pockets $5/6$ and $4$). Such sign reversal behavior is similar to the cases of the iron-based superconductors. Therefore, here the physical origin of the in-gap states is the same with the case in the iron-based superconductor~\cite{zhang,tsai}, which is suggested to be caused by
the Andreev reflection due to the opposite phases of the order
parameters~\cite{jhu}. For the case of the $s$-wave and $f$-wave pairing states, as is seen in Fig.~6(c), the maximum gap is always
far away from the Fermi surface, as a result, the effective superconducting gap magnitudes are much smaller. And
the phases of the gap are in disorder thus no in-gap resonance peaks exist.

\begin{figure}
\centering
\includegraphics[width=0.4\textwidth]{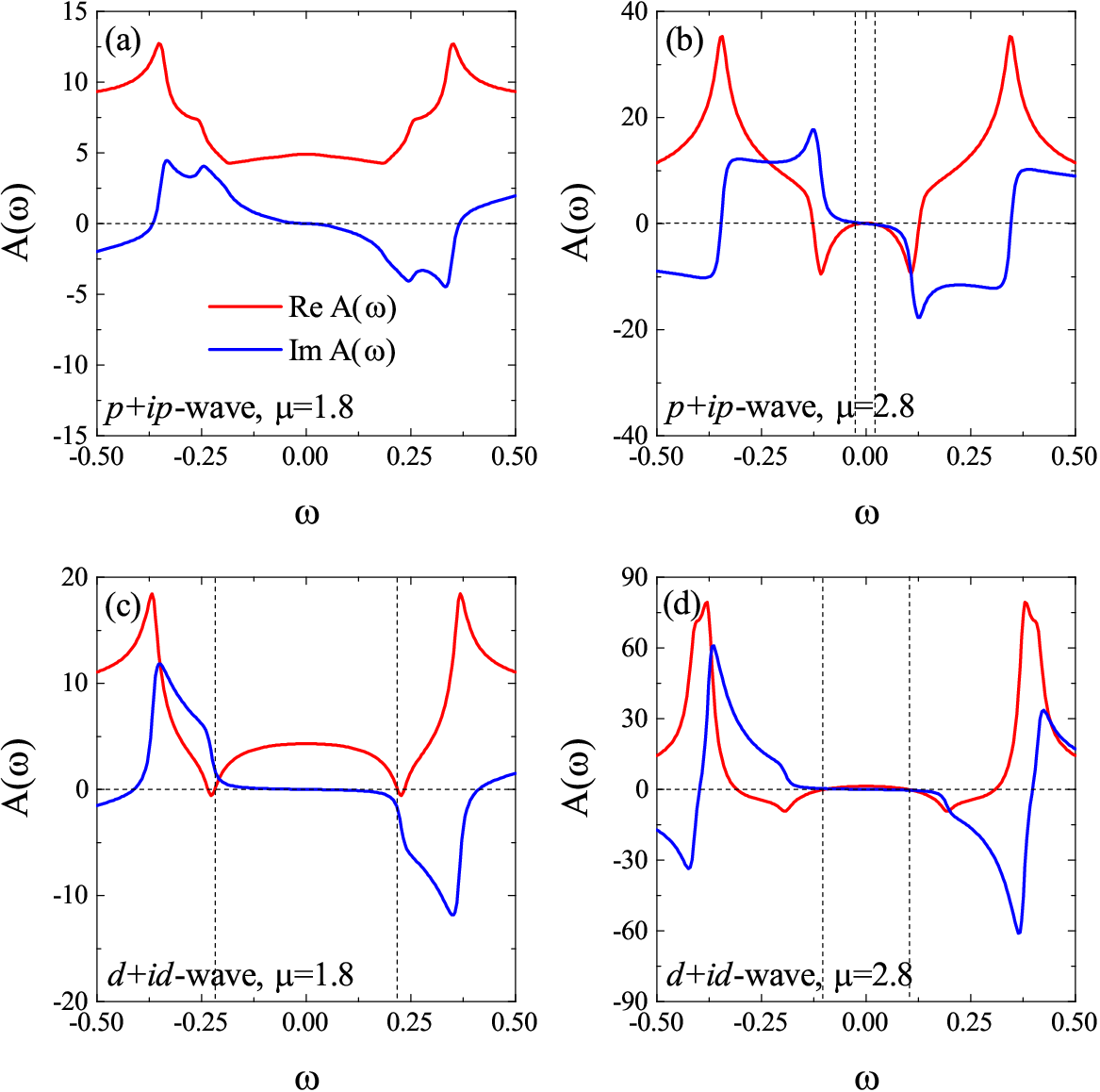}
\caption{(Color online) The real and imaginary parts of the denominator of the $T$-matrix for the $p+ip$-wave and $d+id$-wave pairing symmetries with $\Delta_t = 0.2$ and $V_s=10$.}
\label{fig7}
\end{figure}

The phase analysis and the sign reversal picture can explain qualitatively the emergence of the in-gap bound states. While the sign reversal of the order parameters is usually not the sufficient condition for the bound states.
An alternative method to explain the
the impurity induced resonance peaks can be achieved through discussing the denominator of
 the $T$-matrix [$A(\omega)$] from Eq.\ref{Tmatrix} with $A(\omega)=\mid {\hat I - \hat U_0 \frac{1}{N}\sum\limits_{\bf{k}} {\hat G_0 ({\bf{k}},\omega )} }\mid$, namely, its imaginary part Im$A(\omega)$ at low energies is usually rather small due to the existence of the superconducting gap. Then a resonance occurs when its real part Re$A(\omega)$ approaches to zero at a certain low energy.
The denominator of
 the $T$-matrix with the $p+ip$ and $d+id$ pairing symmetries are presented in Fig.\ref{fig7}. As is seen, its imaginary part at low energies is indeed rather small. Then a resonance occurs
if its real part approaches to zero.
 For the case of the $p+ip$-wave pairing state, as the chemical potential is small ($\mu=1.8$), as is seen in Fig.\ref{fig7}(a), the real part of the denominator does not approach to the zero value for all of the energies we considered. Therefore, at this doping level no in-gap bound states exist. As the chemical potential increases ($\mu=2.8$), [Fig.\ref{fig7}(b)], Re$A(\omega)$ crosses the zero energy axis at two symmetrical low energies, as a result, the in-gap bound states emerge at these two energies.
For the case of the $d+id$-wave pairing state, as is seen in Figs.\ref{fig7}(c) and \ref{fig7}(d), Re$A(\omega)$  cross the zero energy axis for both two chemical potentials we considered. Therefore, the in-gap bound states are robust. Moreover, since the spectra for the $d+id$ pairing is fully gapped and the effective gap magnitudes
are larger, the imaginary parts of the denominator [Im$A(\omega)$] are rather small. As a result, the intensity of the in-gap peaks for the $d+id$ pairing symmetry are rather strong and may be detected easily.

Finally, we discuss whether different pairing states can be resolved from the LDOS near a single impurity. For the cases of the $p+ip$-wave
and $d+id$-wave pairing symmetries, at highly doping level where the Fermi level is close to the Van Hove singularities, there exist in-gap resonance peaks
induced by the impurity, however, the spectra for the $d+id$ pairing symmetry is fully gapped and
the in-gap resonance peaks for the $d+id$-wave pairing are much stronger than that for the $p+ip$-wave pairing.
For the cases of the extended $s$-wave and $f$-wave pairing symmetries, there are no in-gap
states, and the spectra for the $s$-wave pairing symmetry are fully gapped and those for the $f$-wave one are nodal.
Obviously, at low doping levels, the $d+id$-wave pairing symmetry is different from other three
ones, i.e., there exist strong in-gap peaks for the case of $d+id$-wave pairing symmetry. Therefore, we conclude that, the impurity effect indeed provides some useful information and the pairing symmetry may be resolved at some typical doping levels.
Especially, the $d+id$ pairing symmetry may be distinguished from other three pairing symmetries.

\section{summary}

In summary, we have studied theoretically the single impurity effect of graphene-based superconductors. Four different pairing symmetries, i.e., the $p+ip$-wave pairing,
the $d+id$-wave pairing, the extended $s$-wave pairing and the $f$-wave pairing, are considered. Robust in-gap resonance states are revealed for the $d+id$ pairing symmetry.
 For the cases of the $p+ip$-wave pairing, the in-gap
resonsance states are sensitive to the doping level and they exist at the highly doped sample. For the $f$-wave and extended $s$-wave pairing symmetries, no in-gap resonance states are obtained
for all of the parameters we considered. All of the features can be explained through analyzing order parameter along the Fermi pockets and the denominator of the $T$-matrix. We conclude that the impurity effect
may provide useful information to resolve different pairing symmetries in graphene-based superconductors.

\begin{acknowledgments}
This work was supported by the NSFC (Grant No. 12074130) and Science and Technology Program of Guangzhou (Grant No. 2019050001).
\end{acknowledgments}


\begin{thebibliography}{99}
\bibitem{tonn} C. Tonnoir, A. Kimouche, J. Coraux, L. Magaud, B. Delsol, B. Gilles, and C. Chapelier, Phys. Rev. Lett. {\bf 111}, 246805 (2013).
\bibitem{chin} Satoru Ichinokura, Katsuaki Sugawara, Akari Takayama, Takashi Takahashi, and Shuji Hasegawa, Acs Nano {\bf 10}, 2761 (2016).
\bibitem{chap} Satoru Ichinokura, Katsuaki Sugawara, Akari Takayama, Takashi Takahashi, and Shuji Hasegawa, Sci. Rep. {\bf 6}, 23254 (2016).
\bibitem{ludb} B. M. Ludbrook, G. Levy, P. Nigge, M. Zonno, M. Schneider, D. J. Dvorak, C. N. Veenstra, S. Zhdanovich, D. Wong, P. Dosanjh, C. Stra${\beta}$er, A. Stohr,
                S. Forti, C. R. Ast, U. Starke, A. Damascelli, Proc. Natl. Acad. Sci. U.S.A., {\bf 112}, 11795 (2015).
\bibitem{bern} A. Di Bernardo, O. Millo, M. Barbone, H. Alpern, Y. Kalcheim, U. Sassi, A. K. Ott, D. De Fazio, D. Yoon,
                M. Amado, A.C. Ferrari, J. Linder, and J. W. A. Robinson, Nat. Commun. {\bf 8}, 14024 (2017).
\bibitem{ycao} Yuan Cao, Valla Fatemi, Shiang Fang, Kenji Watanabe, Takashi Taniguchi, Efthimios Kaxiras, and Pablo Jarillo-Herrero, Nature {\bf 556}, 43 (2018).
\bibitem{ucho} Bruno Uchoa and A. H. Castro Neto, Phys. Rev. Lett. {\bf 98}, 146801 (2007).
\bibitem{kopn} N. B. Kopnin and E. B. Sonin, Phys. Rev. Lett. {\bf 100}, 246808 (2008).
\bibitem{jacob} Jacob Linder, Annica M. Black-Schaffer, Takehito Yokoyama, Sebastian Doniach, and Asle Sudb${\o}$, Phys. Rev. B {\bf 80}, 094522 (2009).
\bibitem{anni} Annica M. Black-Schaffer and Sebastian Doniach, Phys. Rev. B {\bf 81}, 014517 (2010).
\bibitem{tma} T. Ma, F. Yang, H. Yao, and H. Q. Lin, Phys. Rev. B {\bf 90}, 245114 (2014).
\bibitem{faye} J. P. L. Faye, P. Sahebsara, and D. Senechal, Phys. Rev. B {\bf 92}, 085121 (2015).
\bibitem{nand} Rahul Nandkishore, L. S. Levitov and A. V. Chubukov, Nat. Phys. {\bf 8}, 158 (2012).
\bibitem{kies} Maximilian L. Kiesel, Christian Platt, Werner Hanke, Dmitry A. Abanin, and Ronny Thomale, Phys. Rev. B {\bf 86}, R020507 (2012).
\bibitem{nand1} Rahul Nandkishore, Ronny Thomale, and Andrey V. Chubukov, Phys. Rev. B {\bf 89}, 144501 (2014).
\bibitem{xiao} Long-Yun Xiao, Shun-Li Yu, Wei Wang, Zi-Jian Yao, and Jian-Xin Li, Europhys. Lett. {\bf 115}, 27008 (2016).
\bibitem{hoss} M. V. Hosseini and M. Zareyan, Phys. Rev. Lett. {\bf 108}, 147001 (2012).
\bibitem{lado} J. L. Lado and J. Fernandez-Rossier, 2D Mater. {\bf 3}, 025001 (2016).
\bibitem{zhu} A. V. Balatsky, I. Vekhter, and Jian-Xin Zhu, Rev. Mod. Phys. 78, 373 (2006).
\bibitem{jhu} Chia-Ren Hu, Phys. Rev. Lett. {\bf 72}, 1526 (1994).
\bibitem{zhang} D. G. Zhang, Phys. Rev. Lett. {\bf 103}, 186402 (2009).
\bibitem{tsai} W. F. Tsai, Y. Y. Zhang, C. Fang, and J. P. Hu, Phys. Rev B {\bf 80}, 064513 (2009).
\bibitem{Hudson} E.W. Hudson, S.H. Pan, A.K. Gupta, and K.W. Ng, Science {\bf 88}, 285 (1999).
\bibitem{Morr} D.K. Morr, Phys. Rev. Lett. {\bf 89}, 106401 (2002).
\bibitem{Andrenacci} N. Andrenacci, G.G.N. Angilella, H. Beck, and R. Pucci, Phys. Rev. B {\bf 70}, 024507 (2004).
\bibitem{glad} Oladunjoye A. Awoga and Annica M. Black-Schaffer, Phys. Rev. B {\bf 97}, 214515 (2018).
\bibitem{pell} F.M.D. Pellegrino, G.G.N. Angilella, and R. Pucci, Eur. Phys. J. B {\bf 76}, 469 (2010).
\bibitem{wehl} T. O. Wehling, H. P. Dahal, A. I. Lichtenstein, and A. V. Balatsky, Phys. Rev. B {\bf 78}, 035414 (2008).
\bibitem{Scherer} Michael M. Scherer, Physics {\bf 13}, 161 (2020).
\bibitem{Rosenzweig} P. Rosenzweig, H. Karakachian, D. Marchenko, K. K$\ddot{u}$ster, and U. Starke, Phys. Rev. Lett. {\bf 125}, 176403 (2020).
\bibitem{lot} Tomas Lothman and Annica M. Black-Schaffer, Phys. Rev. B {\bf 90}, 224504 (2014).
\end{thebibliography}
\end{document}